\documentclass[twocolumn,pre,amsmath,floatfix]{revtex4}
\usepackage{graphicx,color}
\newcommand{\figising}{%
\begin{figure}[h]%[htbp]
   \centering
   \includegraphics[width=3.in]{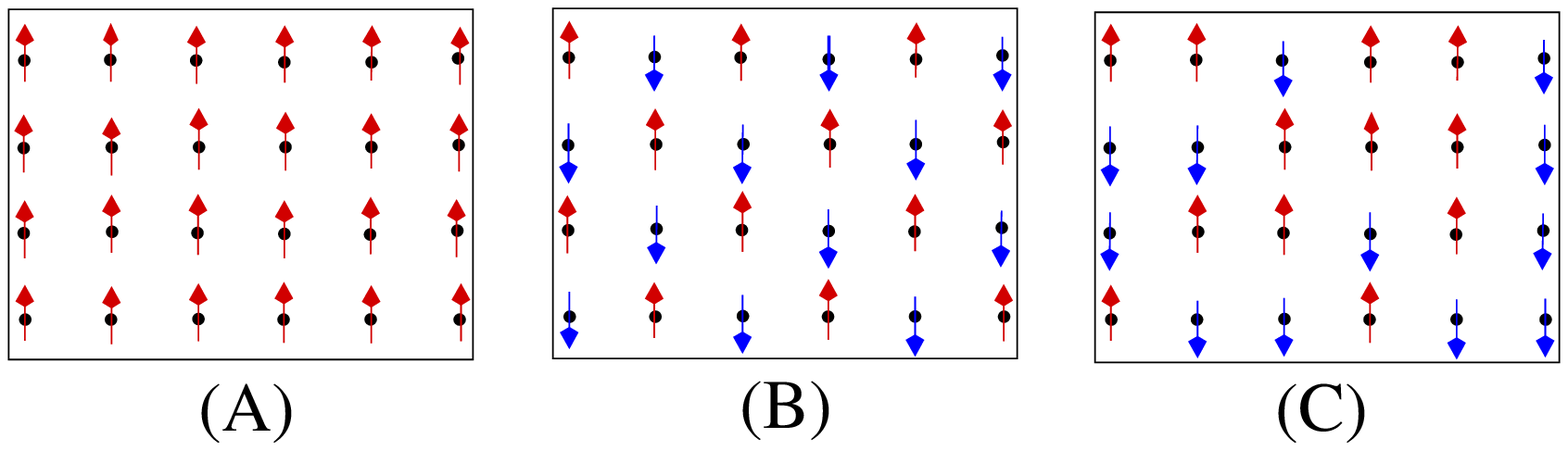}
   \caption{Ising magnet. Spins $\pm 1$ are represented by arrows
     pointing up or down.  (A) A ferromagnetic state, (B) an
     antiferromagnetic state, and (C) a seemingly random configuration.}
   \label{fig:1}
 \end{figure}
}
\newcommand{\figgas}{%
\begin{figure}[h]%[htbp]
  \centering
  \includegraphics[width=1in]{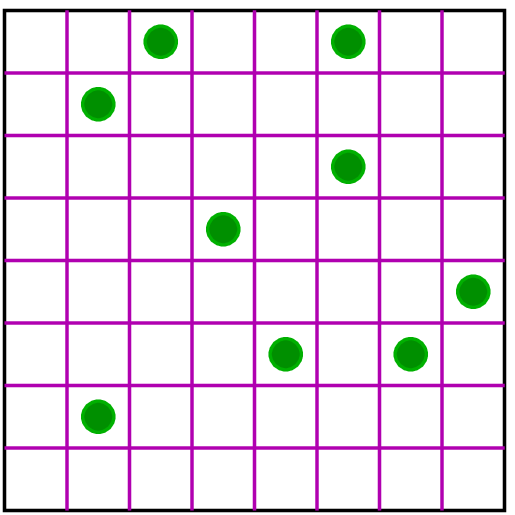}
  \caption{Perfect gas: space divided into cells. The cells are
  occupied by the particles}
  \label{fig:2}
\end{figure}
}
\begin{document}
 
\title{Entropy and perpetual computers}
\author{ Somendra M. Bhattacharjee}
%\email{email: somen@iopb.res.in }
\affiliation{Institute of Physics, Bhubaneswar 751 005, India\\ email:
somen@iopb.res.in }
% Institute of Physics, Bhubaneswar 751 005, India
%email: somen@iopb.res.in }
\begin{abstract}
A definition of entropy via the Kolmogorov algorithmic complexity  is
discussed.  As examples, we show how the meanfield theory for the
Ising model, and the entropy of a perfect gas  can be recovered.  The
connection with computations are pointed out, by paraphrasing the laws
of thermodynamics for computers.   Also discussed is an approach that
may be adopted to develop statistical mechanics using the algorithmic
point of view.
\end{abstract}
\date{\today}   
\maketitle

\section{Introduction}
\label{sec:introduction}

The purpose of this lecture note is to illustrate a  route for the
definition of entropy using our experience with  computers.  In the
process the connection between statistical physics and
computations comes to the fore.

\subsection{What is entropy?}
\label{sec:what-entropy}

This is a question that plagues almost all especially beginning
physics students. There are several correct ways to answer this.
\begin{enumerate}
\item It is the perfect differential that one gets by dividing the
heat transfered by a quantity $T$ that gives us the hot-cold feeling
(i.e. temperature).  ({\color{red}{thermodynamics}})
\item It is the log of the number of states available.
({\color{red}{Boltzmann}})
\item It is something proportional to $-\sum p_i\ln p_i$ where $p_i$
is the probability that the system is in state i.
({\color{red}{Gibbs}})
\item It is just an axiom that there exists an extensive quantity $S$,
obeying certain plausible conditions, from which the usual
thermodynamic rules can be obtained.  ({\color{red}{Callen}})
\end{enumerate}
But  the  colloquial link between  disorder or randomness  and entropy
remains unexpressed though, agreeably,  making a formal connection  is
not easy. Our plan is to establish this missing link {\it a la} Kolmogorov.

Besides these conceptual questions, there is a practical issue that
bugs many who do computer simulations where different configurations
are generated by some set of rules.  In the end one wants to calculate
various thermodynamic quantities which involve both energy and
entropy.  Now, each configuration generated during a simulation or
time evolution has an energy associated with it.  {\it But does it
have an entropy?}  The answer is of course blowing in the wind.  All
thermodynamic behaviours ultimately come from a free energy, say, $F =
\langle E\rangle - TS$ where $E$, the energy, generally known from
mechanical ideas like the Hamiltonian, enters as an average, denoted
by the angular brackets, {\it but no such average for $S$}.  As a
result, one cannot talk of ``free energy'' of a configuration at any
stage of the simulation.  All the definitions mentioned above
associate $S$ to the ensemble, or distributions over the phase space.
They simply forbid the question ``what is the entropy of a
configuration". Too bad!
\subsection{On computers}
\label{sec:computers-1}
Over the years we have seen the size of computers shrinking, speed
increasing and power requirement going down. Centuries ago a question
that tickled scientists was the possibility of converting heat to work
or finding a perfect engine going in a cycle that would completely
convert heat to work. A current version of the same problem would be:
Can we have a computer that does computations but at the end does not
require any energy.  Or, we take a computer, draw power from a
rechargeable battery to do the computation, then do the reverse
operations and give back the energy to the battery.  Such a computer
is in principle a perpetual computer.  {\it Is it possible? }

What we mean by a computer is a machine or an object that implements a
set of instructions without any intelligence. It executes whatever it
has been instructed to do without any decision making at any point.
At the outset, without loss of generality, we choose binary (0,1) as
the alphabet to be used, each letter to be called a bit.  The job of
the computer is to manipulate a given string as per instructions. Just
as in physics, where we are interested in the thermodynamic limit of
infinitely large number of particles, volumes etc, we would be
interested in infinitely long strings. The question therefore is ``can
bit manipulations be done without cost of energy?"
\section{Randomness}
\label{sec:randomness}
The problem that a configuration can not have an entropy has its
origin in the standard statistical problem that a given outcome of an
experiment cannot be tested for randomness.  E.g., one number
generated by a random number generator cannot be tested for
randomness.

For concreteness, let us consider a general model system of a magnet
consisting of spins $s_i=\pm 1$ arranged on a square lattice with $i$
representing a lattice site.  If necessary, we may also use an energy
(or Hamiltonian) $E = -J\sum_{<ij>} s_is_j$ where the sum is over
nearest neighbours (i.e. bonds of the lattice).  Suppose the
temperature is so high that each spin can be in anyone of the two
states $\pm 1$ with equal probability. We may generate such a
configuration by repeated tossing of a fair coin. If we get
$+--++----+-+-+----+ $ ($+$:H,$-$:T) is it a random configuration? Or
Can the configurations of spins as shown in Fig. \ref{fig:1} be
considered random?
 
\figising

With $N$ spins (or bits), under tossing of a fair coin, the
probability of getting Fig. \ref{fig:1}(A) is $2^{-N}$ and so is the
probability of (B) or (C).  Therefore, the fact that a process is
random cannot be used to guarantee randomness of the sequence of
outcomes.  Still, we do have a naive feeling.  All Heads in $N$ coin
toss experiments or strings like 1111111...  (ferro state of Fig.
\ref{fig:1}(A)) or 10101010...  (anti-ferro state of Fig
\ref{fig:1}(B)) are never considered random because one can identify a
pattern, but a string like 110110011100011010001001... (or
configuration of Fig \ref{fig:1}(C)) may be taken as random.  {\it But
what is it that gives us this feeling?}

\subsection{Algorithmic approach}
\label{sec:algorothmic-approach}
The naive expectation can be quantified by a different type of
arguments, not generally emphasized in physics. Suppose I want to
describe the string by a computer programme; or rather by an
algorithm. Of course there is no unique ``programming" language nor
there is ``a" computer - but these are not very serious issues.  We
may choose, arbitrarily, one language and one computer and transform
all other languages to this language (by adding "translators") and
always choose one particular computer. The two strings, the ferro and
the anti-ferro states, can then be obtained as outputs of two very
small programmes,
\begin{verbatim}
(A) Print 1   5 million times (ferro state)
(B) Print 10  2.5 million times (antiferro state)
\end{verbatim}
In contrast, the third string would come from
\begin{verbatim}
(C) Print 110110011100...  (disordered state)
\end{verbatim}
so that the size of the programme is same as the size of the string
itself.  This example shows that the size of the programme gives an
expression to the naive feeling of randomness we have.  We may then
adopt it for a quantitative measure of randomness.
\begin{quote}
  Definition : Let us define {\it randomness} of a string as the size
  of the {\it minimal} programme that generates the string.
\end{quote}
The crucial word is ``minimal".  In computer parlance what we are
trying to achieve is a compression of the string and the minimal
programme is the best compression that can be achieved.

Another name given to what we called ``randomness'' is {\it
  complexity}, and this particular measure is called Kolmogorov
  algorithmic complexity.  The same quantity, randomness, is also
  called information, because the more we can compress a string the
  less is the information content.  Information and randomness are
  then two sides of the same coin: the former expressing a positive
  aspect while the 2nd a negative one!

Let $K({\sf c})$ be a programme for the string of configuration ${\sf
  c}$ and let us denote the length of any string by $\mid ...\mid$.
  The randomness or complexity is
\begin{equation}
  \label{eq:10}
  {\cal S}({\sf c})=\min {\mid}K({\sf c})|.
\end{equation}
We now define {\it a string as random}, if its randomness or
complexity is similar to the length of the string, or, to be
quantitative, if randomness is larger than a pre-chosen threshold,
e.g, say, ${\cal S}({\sf c}) > {\mid}{\sf c}| - 13$.  The choice of
$13$ is surely arbitrary here and any number would do.

\subsubsection{Comments}
\label{sec:comments1}
A few things need to be mentioned here. {\it (i)} By definition, a
minimal programme is random, because its size cannot be reduced
further.  {\it (ii)} It is possible to prove that a string is {\it
not} random by explicitly constructing a small programme, but it is
not possible to prove that a string {\it is} random. This is related
to G\"odel's incompleteness theorem.  For example, the digits of $\pi$
may look random (and believed to be so) until one realizes that these
can be obtained from an efficient routine for, say, $\tan^{-1}$.  We
may not have a well-defined way of constructing minimal algorithms,
but we agree that such an algorithm exists.  {\it (iii)} The
arbitrariness in the choice of language leads to some indefiniteness
in the definition of randomness which can be cured by agreeing to add
a translator programme to all other programmes.  This still leaves the
differences of randomness of two strings to be the same.  In other
words, randomness is defined upto an arbitrary additive constant.
Entropy in classical thermodynamics also has that arbitrariness.  {\it
(iv)} Such a definition of randomness satisfies a type of
subadditivity condition $S({\sf c}_1+{\sf c}_2) \le S({\sf c}_1)+
S({\sf c}_2) +O(1)$, where the $O(1)$ term cannot be ignored.

\subsection{Entropy}
\label{sec:entropy}
Accepting that this Kolmogorovian approach to randomness makes sense
and since we connect randomness in a physical system with entropy, let
us associate this randomness ${\cal S}({\sf c})$ with the entropy of
that string or configuration ${\sf c}$.  For an ensemble of strings or
configurations with probability $p_i$ for the $i$-{th} string or
configuration ${\sf c}_i$, the average entropy will be defined by
\begin{equation}
  \label{eq:2}
 S_{\rm K} = \sum_i p_i  {\cal S}({\sf c}_i)
\end{equation}
(taking the Boltzmann constant $k_B=1$).    We shall claim that this is
the thermodynamic entropy we are familiar with.

Since the definition of entropy in Eq. (\ref{eq:2}) looks ad hoc, let
us first show that this definition gives us back the results we are
familiar with.  To complete the story, we then establish the
equivalence with the Gibbs definition of entropy.

\subsection{Example I: Mean filed theory for the Ising model}
\label{sec:example-i:-mean}
Consider the Ising problem. Let us try to write the free energy of a
state with $n_+$ + spins and $n_ -\ -$ spins with $n_+ + n_-=N$. The
number of such configurations is
\begin{equation}
  \label{eq:3}
  \Omega = \frac{N!} {n_+! ~n_-!}.
\end{equation}
An ordered list (say lexicographical) of all of these $\Omega$
configurations is then made.  If all of these states are equally
likely to occur then one may specify a state by a string that
identifies its location in the list of configurations.  The size of
the programme is then the number of bits required to store numbers of
the order of $\Omega$.  Let $S$ be the number of bits required.  For
general $N, n_+, n_-$, $S$ is given by
\begin{equation}
  \label{eq:4}
  2^S =\Omega \qquad \Longrightarrow\qquad S = \log_2\Omega.
\end{equation}
Stirling's approximation then gives
\begin{eqnarray}
  \label{eq:1}
 S & = &n_+\log_2 n_++   n_-\log_2 n_- \nonumber\\
& =& N ~[p \log_2 p +(1-p) \log_2 (1-p)], 
\end{eqnarray}
with $p = n_+/N$, the probability of a spin being up.  Resemblance of
Eq. (\ref{eq:4}) with the Boltzmann formula for entropy (Sec.
\ref{sec:introduction}) should not go unnoticed here.  Eq.
(\ref{eq:1}) is the celebrated formula that goes under the name of
entropy of mixing for alloys, solutions etc.

\subsubsection{Comments}
\label{sec:comments2}
It is important to note that no attempt has been made for
``minimalizations" of the algorithm or in other words we have not
attempted to compress $\Omega$.  For example, no matter what the
various strings are, all of the N spin configurations can be generated
by a loop (algorithm represented schematically)
\begin{verbatim}
     i = 0
10   i = i+1 
     L = length of i in binary
     Print 0  (N-L) times, then  "i" in binary
     If  ( i < N ) go to 10
     stop
\end{verbatim}
By a suitable choice of $N$ ({\it e.g.}, $N =11.....1$) the code for
representation of $N$ can be shortened enormously by compressing $N$.
This shows that one may generate all the spin configurations by a
small programme though there are several configurations that would
require individually much bigger programmes.  This should not be
considered a contradiction because it produces much more than we want.
It is fair to put a restriction that the programmes we want should be
self delimiting (meaning it should stop without intervention) and
should produce just what we want, preferably no extra output. Such a
restriction then automatically excludes the above loop.

Secondly, many of the numbers in the sequence from $1$ to $\Omega$ can
be compressed enormously. However, what enumeration scheme we use,
cannot be crucial for physical properties of a magnet, and therefore,
we do need $S$ bits to convey an arbitrary configuration.  It is also
reassuring to realize that there are random (i.e.  incompressible)
strings in $2^N$ possible $N$-bit strings. The proof goes as follows.
If an $N$-bit string is compressible, then the compressed length would
be $\leq N-1$.  But there are only $2^{N-1}$ such strings.  Now the
compression procedure has to be one to one (unique) or otherwise
decompression will not be possible.  Hence, for every $N$, there are
strings which are not compressible and therefore random.

  A related question is the time required to run a programme.  What we
have defined so far is the ``space'' requirement.  It is also possible
to define a ``time complexity'' defined by the time required to get
the output.  In this note we avoid this issue of time altogether.

\subsubsection{Free energy}
\label{sec:free-energy}
In the Kolmogorov approach we can now write the free energy of any
configuration, ${\sf c}_i$ as $F_i=E_i - TS_i$ with the thermodynamic
free energy coming from the average over all configurations,
$$F\equiv\langle F\rangle=\langle E\rangle-T\langle S\rangle .$$ If we
now claim that $S$ obtained in Eq. ~(\ref{eq:1}) is the entropy of any
configuration, and since no compression is used, it is the same for
all (this is obviously an approximation), we may use $\langle S\rangle
= S$.  The average energy may be approximated by assuming random
mixture of up and down spins with an average value $\langle
s\rangle=p-(1-p)$.  If $q$ is the number of nearest neighbours ($4$
for a square lattice), the free energy is then given by
\begin{equation}
  \label{eq:5}
\frac{F}{N} = \frac{q}{2}J (2p-1)^2~ -~ T [p \log p + (1-p)\log
(1-p)].
\end{equation}
Note that we have not used the Boltzmann or the Gibbs formula for
entropy.  By using the Kolmogorov definition what we get back is the
mean field (or Bragg-Williams) approximation for the Ising model.  As
is well-known, this equation on minimization of $F$ with respect to
$p$, gives us the Curie-Weiss law for magnetic susceptibility at the
ferro-magnetic transition.  No need to go into details of that because
the purpose of this exercise is to show that the Kolmogorov approach
works.

\subsection{Example II: Perfect gas}
\label{sec:example-ii:-perfect}
A more elementary example is the S\"ackur-Tetrode formula for entropy
of a perfect gas.  We use cells of small sizes $\Delta V$ such that
each cell may contain at most one particle. For N particles we need
$\Omega =(V/\Delta V)^{N}$ numbers to specify a configuration, because
each particle can be in one of $V/\Delta V$ cells. The size in bits is
$S = N \log_2 \frac{V}{\Delta V }$ so that the change in randomness or
entropy as the volume is changed from $V_i$ to $V_f$ is
\begin{equation}
  \label{eq:6}
 \Delta S = N \log_2 \frac{V_f}{V_i}.
\end{equation}
The indistinguishability factor can also be taken into account in the
above argument, but since it does not affect Eq. (\ref{eq:6}), we do
not go into that.  Similarly momentum contribution can also be
considered.

\figgas

It may be noted here that the work done in isothermal expansion of a
perfect gas is
\begin{equation}
  \label{eq:7} \int_{V_{i}}^{V_{f}} P\ dV = N k_{\rm B}T \ln
\frac{V_f}{V_i} = (k_{\rm B} \ln 2) T \Delta S.
\end{equation}
Where $P$ is the pressure satisfying $PV=Nk_{\rm B}T$ and $\Delta S$
is defined in Eq. (\ref{eq:6}).  Both Eqs. (\ref{eq:6}) and
(\ref{eq:7}) are identical to what we get from thermodynamics.  The
emergence of $\ln 2$ is because of the change in base from $2$ to $e$.

It seems logical enough to take this route to the definition of
entropy and it would remove much of the mist surrounding entropy in
the beginning years of a physics student.

\section{Computers}
\label{sec:computers}
\subsection{On computation}
For the computer problem mentioned in the Introduction, one needs to
ponder a bit about reality.  In thermodynamics, one considers a
reversible engine which may not be practical, may not even be
implementable.  But a reversible system without dissipation can always
be justified.  Can one do so for computers?
\subsubsection{Reversible computers?}
To implement an algorithm (as given to it), one needs logic circuits
consisting of say AND and NAND gates (all others can be built with
these two) each of which requires two inputs (a,b) to give one output
(c). By construction, such gates are irreversible: given c, one can
not reconstruct a and b. However it is possible, at the cost of extra
signals, to construct a reversible gate (called a Toffoli gate) that
gives AND or NAND depending on a third extra signal. The truth table
is given in Appendix \ref{sec:toffoli-gate}. Reversibility is
obvious. A computer based on such reversible gates can run both ways
and therefore, after the end of manipulations, can be run backwards
because the hardware now allows that.  Just like a reversible engine,
we now have a reversible computer. All our references to computers
will be to such reversible computers.

\subsubsection{Laws of computation}
Let us try to formulate a few basic principles applicable to
computers. These are rephrased versions of laws familiar to us.

\begin{quote}
  {\bf Law I}:  It is not possible to have perpetual computation.
\end{quote}
In other words, we cannot have a computer that can read a set of
instructions and carry out computations to give us the output {\it
without any energy requirement}.  Proving this is not straight forward
but this is not inconsistent with our intuitive ideas.  We won't
pursue this.  This type of computer may be called perpetual computer
of type I.  First law actually forbids such perpetual computers.

\begin{quote}
  {\bf Law II}: It is not possible to have a computer whose sole
  purpose is to draw energy from a reversible source, execute the
  instructions to give the output and run backward to deliver the
  energy back to source, and yet leave the memory at the end in the
  original starting state.
\end{quote}
A computer that can actually do this will be called a perpetual
computer of second kind or type II.

\subsubsection{What generates heat?}
In order to see the importance of the second law, we need to consider
various manipulations on a file (which is actually 
a string).  Our interest is in long strings (length going to infinity
as in thermodynamic limit in physics).  Now suppose we want to edit
the file and change one character, say, in the 21st position.  We may
then start with the original file and add an instruction to go to that
position and change the character.  As a result the edit operation is
described by a programme which is almost of the same length (at least
in the limit of long strings) as the original programme giving the
string.  Therefore there is no change in entropy in this editing
process.  Suppose we want to copy a file.  We may attach the copy
programme with the file.  The copy programme itself is of small size.
The copy process therefore again does not change the entropy.  One may
continue with all the possible manipulations on a string and convince
oneself that all (but one) can be performed at constant entropy.

The exceptional process is {\it delete or removal of a file}.  There
is no need of elaboration that this is a vital process in any
computation.  When we remove a file, we are replacing the entire
string by all zeros - a state with negligible entropy.  It is this
process that would reduce the entropy by $N$ for $N$ characters so
that in conventional units the heat produced at temperature $T$ is
$Nk_{\rm B}T \ln 2$ (see Eq. (\ref{eq:7})).  We know from physics that
entropy reduction does not happen naturally (we cannot cool a system
easily).

\subsubsection{Memory as fuel}
We can have a reversible computer that starts by taking energy from a
source to carry out the operations but to run it backward (via Toffoli
gates) it has to store many redundant information in memory.  Even
though the processes are iso-entropic and can be reversed after
getting the output to give back the energy to the source, we {\bf no
longer} have the memory in the same ``blank'' state we started with.
To get back to that ``blank'' state, we have to clear the memory
(remove the strings).  This last step lowers the entropy, a process
that cannot be carried out without help from outside. If we do not
want to clear the memory, the computer will stop working once the
memory is full.

This is the second law that prohibits perpetual computer of second
kind.  The similarity with thermodynamic rules is apparent.  To
complete the analogy, a computer is like an ``engine'' and memory is
the fuel.  From a practical point of view, this loss of entropy is
given out as heat (similar to latent heat on freezing of water).
Landauer in 1961 pointed out that the heat produced due to this loss
of entropy is $k_{\rm B} T \ln 2$ per bit or $N k_{\rm B} T \ln 2$ for
$N$ bits.  For comparison, one may note that $N k_{\rm B} \ln 2$ is
the total amount of entropy lost when an Ising ferromagnet is cooled
from a very high temperature paramagnetic phase to a very low
temperature ferromagnetic phase.  If the process of deletion on a
computer occurs very fast in a very small region of space, this heat
generation can create problem.  It therefore puts a limit on
miniaturization or speed of computation.  Admittedly this limit is not
too realistic because other real life processes would play major roles
in determining speed and size of a computer. See Appendix
\ref{sec:heat-generated-chip} for an estimate of heat generated.

\subsection{Communication}
\label{sec:communication}
\subsubsection{The problem}
Let us now look at another aspect of computers namely transmission of
strings (or files) or communication.  This topic actually predates
computers.  To be concrete, let us consider a case where we want to
transmit images discretized into small cells of four colours,
{\color{black}{R}}{\color{black}{G}}{\color{black}{B}}{\color{black}{Y}}
with probabilities
$$p({\rm R})=1/2, p({\rm G})=1/4, p({\rm B})=p({\rm Y})=1/8.$$  
The question in communication is: ``What is the minimal length of string
(in bits) required to transmit any such image?''  

\subsubsection{Kolmogorov and Shannon's theorem}
There are two possible ways to answer this question. The first is
given by the Kolmogorov entropy (= randomness = complexity) while the
second is given by a different powerful theorem called Shannon's
noiseless coding theorem.  Given a long string ${\cal C}_j$ of say
${N}$ characters, if we know its Kolmogorov entropy ${\cal S}_j$ then
that has to be the smallest size for that string. If we now consider
all possible ${N}$ character strings with ${\cal P}_j$ as the
probability of the $j$th string, then $S_{\rm K}=\sum_j {\cal P}_j
{\cal S}_j$ is the average number we are looking for.  Unfortunately
it is not possible to compute ${\cal S}_j$ for all cases.  Here we get
help from Shannon's theorem.  The possibility of transmitting a signal
that can be decoded uniquely is guaranteed with probability 1, if the
average number of bits per character $= -\sum p_i \log_2 p_i$ where
$p_i$'s are the probabilities of individual characters.  A proof of
this theorem is given in Appendix \ref{sec:proof-shann-theor}.  Since
the two refer to the same object, they are the same with probability
1, {\it i.e.},
$$S_{\rm K} = - { N} \sum p_i \log_2 p_i.$$

\subsubsection{Examples}
The applicability of the Shannon theorem is now shown for the
above example.  To choose a coding scheme, we need to restrict
ourselves to {\it prefix} codes (i.e. codes that do not use one code
as the ``prefix'' of another code.  As an example, if we choose ${\rm
{\color{black}{R}}}\equiv 0, {\rm {\color{black}{G}}}\equiv 1, {\rm
{\color{black}{B}}}\equiv 10, {\rm {\color{black}{Y}}}\equiv 11$,
decoding cannot be unique. E.g. what is 010?
{\color{black}{R}}{\color{black}{G}}{\color{black}{R}} or
{\color{black}{R}}{\color{black}{B}}?  Nonuniqueness here came from
the fact that {\color{black}{B}} ($10$) has the code of
{\color{black}{G}} ($1$) as the first string or prefix.  A scheme
which is prefix free is to be called a prefix code.

For our original example, we may choose ${\rm
  {\color{black}{R}}}\equiv 0, {\rm {\color{black}{G}}}\equiv 10,{\rm
  {\color{black}{B}}}\equiv 110, {\rm {\color{black}{Y}}}\equiv 111$
as a possible coding scheme to find that the average length required
to transmit a colour is
\begin{equation}
  \label{eq:8}
 \langle l\rangle \equiv 1\times \frac{1}{2} + 2 \times \frac{1}{4} +
2\times 3\times \frac{1}{8}=\frac{7}{4}.
\end{equation}
It is a simple exercise to show that any other method would only
increase the average size.  What is remarkable is that 
 $$-\sum_i p_i\log_2 p_i = 7/4,$$ 
an expression we are familiar with
from the Gibbs entropy and also see in the Shannon theorem.

In case the source changes its pattern and starts sending signals with
equal probability
$$p({\rm R})= p({\rm G})= p({\rm B})=p({\rm Y})=1/4,$$ 
we may adopt a different scheme with 
$${\rm {\color{black}{R}}}\equiv 00, {\rm {\color{black}{G}}}\equiv
10,{\rm {\color{black}{B}}}\equiv 01, {\rm {\color{black}{Y}}}\equiv
11,$$ 
for which the average length is
$$\langle l\rangle =2 = -\sum_i p_i \log_2 p_i.$$ This is less than
what we would get if we stick to the first scheme.  Such simple
schemes may not work for arbitrary cases as, e.g., for
$$p({\rm R})=1/2, p({\rm G})=\frac{1}{4}-2\epsilon,
p({\rm B})=p({\rm Y})=\frac{1}{8}+\epsilon.$$
In the first scheme we get $\langle l\rangle = \frac{7}{4} + 2\epsilon$
while the second scheme would give $\langle l\rangle =2$.  
In the limit of $\epsilon=1/8$, we can opt for a simpler code 
$${\rm {\color{black}{R}}}\equiv 0,{\rm {\color{black}{B}}}\equiv 10,
{\rm {\color{black}{Y}}}\equiv 11,\quad {\rm with}\quad \langle
l\rangle =3/2.$$ 
One way to reduce this length is then to make a list
of all possible $2^{NS}$ strings, where $S=-\sum p \log_2 p$ in some
particular order and then transmit the item number of the message.
This cannot require more than $S$ bits per character.  We see the
importance of the Gibbs formula but it is called the Shannon entropy.

\subsubsection{Entropy}
It is to be noted that the Shannon theorem looks at the ensemble and
not at each string independently.  Therefore the Shannon entropy
$S=-\sum_i p_i \ln p_i$ is ensemble based, but as the examples of
magnet or noninteracting gas showed, this entropy can be used to get
the entropy of individual strings.

Given a set, like the colours in the above example, we can have
different probability distributions for the elements.  The Shannon
entropy would be determined by that distribution.  In the Kolmogorov
case, we are assigning an ``entropy'' ${\cal S}_j$ to the $j$th long
string or state but  $S_{\rm K}$ is determined by the
probabilities ${\cal P}_j$'s of the long strings which are in turn
determined by the $p$'s of the individual characters.  Since both
refer to the best compression on the average, they have to be
equivalent.  It should however be noted that this equivalence is only
in the limit and is a probability 1 statement meaning that there are
configurations which are almost not likely to occur and they are not
counted in the Shannon entropy.  Instead of the full list to represent
all the configurations (as we did in Eqs. (\ref{eq:3}) and
(\ref{eq:4})), it suffices to consider a smaller list consisting of
the relevant or typical configurations.  They are $2^{-{ N}\sum
p\log_2p}$ in number (see Appendix \ref{sec:proof-shann-theor} for
details), typically requiring $S$ bits per character.  A physical
example may illustrate this.  Even though all configuration of
molecules in a gas are allowed and should be taken into account, it is
known that not much harm is done by excluding those configurations
where all the molecules are confined in a small volume in one corner
of a room.  In fact giving equal weightage to all the configurations
in Eq. (\ref{eq:4}) is one of the sources of approximations of
meanfield theory.

\section{Statistical mechanics}
\label{sec:stat-mech}
We now try to argue that statistical mechanics can also be developed
with the above entropy picture.  To do so, we consider the
conventional canonical ensemble, i.e., a system defined by a
Hamiltonian or energy $H$ in contact with a reservoir or bath with
which it can exchange only energy.  In equilibrium, there is no net
flow of energy from one to the other but there is exchange of energy
going on so that our system goes through all the available states in
phase space.  This process is conventionally described by appropriate
equations of motions but, though not done generally, one may think of
the exchange as a communication problem.  In equilibrium, the system
is in all possible states with probability $p_i$ for the $i$th state
and is always in communication with the reservoir about its
configuration.  The communication is therefore a long string of the
states of the system each occurring independently and identically
distributed (that's the meaning of equilibrium).  It seems natural to
make the hypothesis that nature picks the optimal way of
communication.  We of course assume that the communication is
noiseless.  The approach to equilibrium is just the search for the
optimal communication.  While the approach process has a time
dependence where the ``time'' complexity would play a role, it has no
bearing in equilibrium and need not worry us.  With that in mind, we
may make the following postulates:
\begin{quote}
  (1) In equilibrium, the energy $\langle E \rangle = \sum _i p_i E_i$
   remains constant.\\
 (2) The communication with the reservoir is optimal with entropy
 $S=-\sum_i p_i \ln p_i$ .\\
(3) For a given average energy, the entropy is maximum to minimize
failures in communication.
\end{quote}
The third postulate actually assures that the maximum possible number
of configurations ($=2^S$) are taken into account in the communication
process.  No attempt has been made to see if these postulates can be
further minimized.

With these sensible postulates, we have the problem of maximizing $S$
with respect to $p_i$'s keeping $\langle E\rangle$=constant and
$\sum_i p_i =1$.  A straight forward variational calculation shows
that $p_i = \exp(-\beta E_i)/Z$ with $Z=\sum \exp(-\beta E_i)$ being
the standard partition function.  The parameter $\beta$ is to be
chosen properly such that one gets back the average energy.  The usual
arguments of statistical mechanics can now be used to identify $\beta$
with the inverse temperature of the reservoir.

\section{Summary}
\label{sec:summary}

We have tried to show how the Kolmogorov approach to randomness may be
fruitfully used to define entropy and also to formulate statistical
mechanics.  Once the equivalence with conventional approach is
established, all calculations can then be done in the existing
framework.  What is gained is a conceptual framework which lends
itself to exploitation in understanding basic issues of computations.
This would not have been possible in the existing framework.  This
also opens up the possibility of replacing ``engines'' by
``computers'' in teaching of thermodynamics.

\begin{center}
{\bf{Acknowledgments}}
\end{center}

This is based on the C. K. Majumdar memorial talks given in Kolkata on
22nd and 23rd May 2003.  I was fortunate enough to have a researcher
like Prof. Chanchal Kumar Majumdar as a teacher in Science College.  I
thank the CKM Memorial Trust for organizing the memorial talk in
Science College, Kolkata.

\appendix

\section{Toffoli gate}
\label{sec:toffoli-gate}
The truth table of the Toffoli gate is given below.  With three inputs
a,b,c, the output in c$^{\prime}$ is the AND or NAND operation of a
and b depending on c=0 or 1.
\vspace{.25cm}

\begin{minipage}[t]{1.5in}
\includegraphics[width=1.5in]{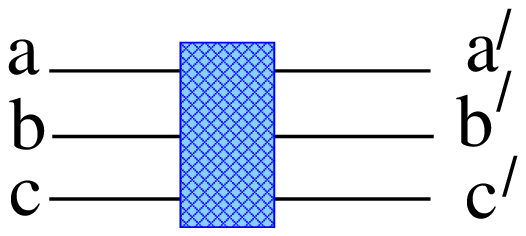}
Fig. A1: Toffoli gate
\end{minipage}
\begin{minipage}[t]{1.5in}
\begin{center}
\begin{tabular}{|ccc|ccc|}
\hline
\ a \ & \ b \ & \ c \ & \ a$^\prime$\ &\ b$^\prime$\ &
\ c$^\prime$\ \\  
\hline
0&0&0&0&0&0\\
0&1&0&0&1&0\\
1&0&0&1&0&0\\
1&1&0&1&1&1\\
\hline
0&0&1&0&0&1\\
0&1&1&0&1&1\\
1&0&1&1&0&1\\
1&1&1&1&1&0\\
\hline
\end{tabular}
\end{center}
\end{minipage}

\section{Proof of Shannon's theorem}
\label{sec:proof-shann-theor}
The  statement of Shannon's noiseless coding theorem is :
\begin{quote}
  If $\langle l\rangle$ is the minimal average code length of an
  optimal code, then
$$S \le  \langle l\rangle \le S +1$$
where $S=-\sum_j p_j\log_2 p_j$.
\end{quote}
The adjective ``noiseless'' is meant to remind us that there is no
error in communication.  A more verbose statement would be 
\begin{quote}
  If we use $N\langle l\rangle$ bits to represent strings of $N$
  characters with Shannon entropy $S$, then a reliable compression
  scheme exists if $\langle l \rangle > S$.  Conversely, if $ \langle
  l\rangle <S$, no compression scheme is reliable.
\end{quote}
The equivalence of the two statements can be seen by recognizing that
$S$ need not be an integer but $\langle l\rangle$ better be.

\subsection{Simple motivation}
\label{sec:simple-motivation}
Let us first go through a  heuristic argument to motivate Shannon's
coding theorem.   Suppose a source is emitting signals $\{{\sf c}_i\}$
independently and identically distributed with two possible values 
${\sf c}_i = 0$  with probability $p_1=p$,  and ${\sf c}_i = 1$  with
probability $p_2=1-p$.   For a long enough string
${\cal C}\equiv {\sf c}_1{\sf c}_2{\sf c}_3{\sf c}_4...{\sf c}_N$
the probability is 
\begin{subequations}
\begin{eqnarray}
  \label{eq:11}
 {\cal P}({\cal C}) 
   &=& p({\sf c}_1)p({\sf c}_2)p({\sf c}_3)p({\sf c}_4)...p({\sf
     c}_N) \qquad \ \ \\
\label{eq:11b}   &\approx& p^{Np}(1-p)^{N(1-p)}\\
    &=& 2^{-N [p\log_2 p + (1-p) \log_2 (1-p)]},
\end{eqnarray}
\end{subequations}
because for large $N$ the number of expected $0$ is $Np$ and $1$ is
$N(1-p)$.  This expression shows that the probability of a long string
is determined by
\begin{equation}
  \label{eq:13}
  S(\{p_j\}) =  -[p \log_2 p + (1-p) \log_2 (1-p)],
\end{equation}
the ``entropy'' for this particular problem.  Note the subtle change
from Eq. (\ref{eq:11}) to Eq. (\ref{eq:11b}).  This use of expectation
values for large $N$ led to the result that most of the strings, may
be called the ``typical'' strings, belong to a subset of $2^{NS}$
strings (out of total $2^N$ strings).

\subsection{What is ``Typical"?}
\label{sec:what-typical}
Let us define a typical string more precisely for any distribution.  A
string of $N$ symbols ${\cal C}={\sf c}_1{\sf c}_2{\sf c}_3{\sf
c}_4...{\sf c}_N$ will be {\it called typical} (or better
$\epsilon$-typical) if
\begin{equation}
  \label{eq:12}
  2^{-N(S+\epsilon)} \le  {\cal P}({\cal C}) \le 2^{-N(S-\epsilon)},
\end{equation}
for any given $\epsilon>0$.
Eq. (\ref{eq:12}) may also be rewritten as
\begin{equation}
  \label{eq:16}
  -\epsilon\le  \left[- N^{-1}\log_2 {\cal P}({\cal C})\right ] -
S \le \epsilon 
\end{equation}

\subsection{How many typical strings?}
\label{sec:how-many-typical}
Now, for random variables ${\sf c}_i$, $X_i$'s, defined by
$X_i=-\log_2 p({\sf c}_i)$, are also independent identically
distributed random variables.  It is then expected that ${\bar
X}=\frac{1}{N}\sum_i X_i$, the average value of $X_i$'s, averaged over
the string for large $N$, should approach the ensemble average,
namely, $\langle X\rangle= -\sum_j p_j \log_2 p_j = S$.  This
expectation comes from the law of large numbers that
\begin{equation}
  \label{eq:14}
  {\rm Prob}\big[~\big| N^{-1}{\sum_i - \log_2 p({\sf  c}_i)} - S
    \big|<\epsilon\big ]
  {\buildrel{N\rightarrow\infty}\over\longrightarrow} \ 1,
\end{equation}
for any $\epsilon >0$.  This means that given an $\epsilon$ we may
find a $\delta>0$ so that the above probability in Eq. \ref{eq:14} is
greater than $1-\delta$.  Recognizing that
\begin{equation}
  \label{eq:15} \sum_i \log_2 p({\sf c}_i) = \log_2 \prod_i p({\sf
 c}_i) = \log_2 {\cal P}({\cal C}),
 \end{equation} Eq. (\ref{eq:14}) implies
\begin{equation}
  \label{eq:17}
  {\rm Prob}\big [~\big|N^{-1} {-\log_2 {\cal P}({\cal C})} -
  S\big| < \epsilon\big] \ge 1 - \delta. 
\end{equation}
We conclude that the probability that a string is typical as defined
in Eqs. (\ref{eq:12}) and (\ref{eq:16}) is $1-\delta$.

Let us now try to estimate the number ${\cal N}_{\rm typ}$, the total
number of typical strings. Let us use a subscript $\mu$ for the
typical strings with $\mu$ going from $1$ to ${\cal N}_{\rm typ}$.
The sum of probabilities ${\cal P}_{\mu}$'s of the typical strings has
to be less than or equal to one, and using the definition of
Eq. (\ref{eq:12}), we have one inequality
\begin{equation}
  \label{eq:18}
  1\ge \sum_{\mu} {\cal P}_\mu \ge \sum_{\mu} 2^{-N(S+\epsilon)} = {\cal
    N}_{\rm typ} 2^{-N(S+\epsilon)}.
\end{equation}
This gives ${\cal N}_{\rm typ} \le 2^{N(S+\epsilon)}$.

Let us now get a lower bound for ${\cal N}_{\rm typ}$.  We have just
established that the probability for a string to be typical is
$1-\delta$.  Using the other limit from Eq. (\ref{eq:12}) we have
\begin{equation}
  \label{eq:19} 1-\delta\le \sum_{\mu} {\cal P}_\mu \le \sum_{\mu}
   2^{-N(S-\epsilon)} = {\cal N}_{\rm typ} 2^{-N(S-\epsilon)},
\end{equation}
which gives ${\cal N}_{\rm typ} \ge (1-\delta)2^{N(S-\epsilon)}$.  The
final result is that the total number of typical strings satisfies 
$2^{N(S+\epsilon)}\ge {\cal N}_{\rm typ} \ge (1-\delta)
2^{N(S-\epsilon)} $ where $\delta >0$ can be chosen small for large
$N$. Hence, in the limit
\begin{equation}
  \label{eq:20}
{\cal N}_{\rm typ}\approx 2^{NS}.
\end{equation}

\subsection{Coding scheme}
\label{sec:coding-scheme}
Now let us choose a coding scheme that requires $Nl$ number of bits
for the string of $N$ characters.  Our aim is to convert a string to a
bit string and decode it - the whole process has to be unique.
Representing the coding and decoding by ``operators'' ${\cal C}$ and
${\cal D}$ respectively, and any string by $\langle {\sf c} |$, what
we want can be written in a familiar form
\begin{eqnarray*}
\label{eq:appb1}
\langle {\sf c}|{\cal  C}|{\cal D}& =& \langle {\sf c}| 
\quad {\rm  for\  all\ } \langle{\sf c}|,\\
{\rm {\tt cat\  myfile|gzip|gunzip} }\ &gives& \ {\rm {\tt myfile}}
\end{eqnarray*}
where the last line is the equivalent ``pipeline'' in a 
 {\tt UNIX} or  {\tt GNU/Linux} system.

Let's take $l > S$.  We may choose an $\epsilon$ such that
$l>S+\epsilon$.  It is a trivial result that $ {\cal N}_{\rm typ} \le
2^{N(S+\epsilon)} < 2^{Nl}$.  Here $2^{Nl}$ is the total number of
possible bit strings.  Hence all the typical strings can be encoded.
Nontypical strings occur very rarely but still they may be encoded.

If $l <S$, then ${\cal N}_{\rm typ} >2^{Nl}$ and obviously all the
typical strings cannot be encoded.  Hence no coding is possible.

This completes the proof of the theorem.

\section{Heat generated in a chip}
\label{sec:heat-generated-chip}
As per a report of 1988, the energy dissipation per logic operation
has gone down from $10^{-3}$ joule in 1945 to $10^{-13}$ joule in
1980's. (Ref: R. W. Keyes, IBM J. Res. Devel. {\bf 32}, 24 (1988) URL:
http://www.research.ibm.com/journal/rd/441/keyes.pdf) For comparison,
thermal energy $k_B T$ at room temperature is of the order of
$10^{-20}$ joule.

If one can pack $10^{18}$ logic gates in one cc operating at
$1$ gigahertz with minimal dissipation of $k_BT$, it  would release
$3$ megawatts of energy.  Can one cool that?
 
A more recent example.  For a pentium 4 at 1.6GHz, if the cpu fan
(that cools the CPU) is kept off, then during operations the cpu
temperature may reach $107C$ (yes Celsius) as monitored by standard
system softwares on an HCL made PC (used for preparation of this
paper).

\end{document}